\documentclass[aps,twocolumn,preprintnumbers,pra,superscriptaddress]{revtex4}

\usepackage{amsmath}
\usepackage{amssymb} 
\usepackage{graphicx} 

\newcommand{\arctanh}[1]{\operatorname{arctan}}

\begin{document}

\title{Simple  non--empirical procedure for
spin--component--scaled MP2 methods applied
to the calculation of dissociation energy curve of
noncovalently--interacting systems}

\author{Ireneusz Grabowski}
\affiliation{Institute of Physics, Faculty of Physics,
Astronomy and Informatics, Nicolaus Copernicus University, Grudziadzka 5,
87-100 Torun, Poland}
\author{Eduardo Fabiano}
\affiliation{National Nanotechnology Laboratory (NNL), Istituto di Nanoscienze-CNR,
Via per Arnesano 16, I-73100 Lecce, Italy }
\author{Fabio Della Sala}
\affiliation{National Nanotechnology Laboratory (NNL), Istituto di Nanoscienze-CNR,
Via per Arnesano 16, I-73100 Lecce, Italy }
\affiliation{Center for Biomolecular Nanotechnologies @UNILE, Istituto Italiano di Tecnologia, 
Via Barsanti, I-73010 Arnesano, Italy}

\date{\today}

\begin{abstract}
We present a
simple and non--empirical method to determine optimal
scaling coefficients, within the (spin--component)--scaled MP2 approach,
for calculating intermolecular potential energies of
noncovalently--interacting systems. The method is based on an
observed proportionality between  (spin--component) MP2 and CCSD(T) energies
for a wide range of intermolecular distances
and allows
to compute with high accuracy a large portion of the dissociation
curve at the cost of a single CCSD(T) calculation.
The
accuracy of the present procedure is assessed for a series of 
noncovalently--interacting test systems: the obtained results
reproduce CCSD(T) quality in all cases and definitely outperform
conventional MP2, CCSD and SCS--MP2 results.
The difficult case of the Beryllium dimer is  also considered.
\end{abstract}

\maketitle

\section{Introduction}

Noncovalent interactions of organic molecules play a fundamental role in
biochemistry, solvation, surface science, and supermolecular chemistry.
An accurate description of the noncovalent interactions is thus very
important for many applications \cite{hobza_book}. 
Actually, the ``golden standard'' for the simulation of noncovalently
interacting complexes is the coupled cluster single and double with 
perturbative triple (CCSD(T)) approach \cite{ragh_CPL_ccsdt}.
However, the CCSD(T) method has a high computational cost, 
scaling as $O(N^7)$, and it is not easily applicable in general, especially when
numerous single--point calculations are required as in the determination
of a potential energy  surface (PES) or for geometry optimizations.

In various applications is thus necessary to recover to lower--level 
computational methods. However, most approaches (including density 
functional theory (DFT) and standard M\o ller--Plesset 
second--order perturbation theory (MP2),
are not fully satisfactory for an accurate description of noncovalent 
interactions and the corresponding PESs
\cite{hobza12,zhao05,zhao06,hobza:2010:review,riley12}. 
Therefore, in the last years different computational schemes, mainly based on
variants of MP2, have been proposed to treat noncovalent interactions 
with sufficient accuracy and relative small computational effort
\cite{hobza12,hobza:2010:review,riley12,marshall11}.

One of such methods is the so called spin--component--scaled MP2 (SCS--MP2)
method \cite{grimme03,WCMS12} which is based on the spin resolved MP2 formula
for the correlation energy
\begin{eqnarray}
\label{eq:scs}
E_c^{SCS-MP2} &=& c_{OS} E^{MP2(OS)} + c_{SS} E^{MP2(SS)} \nonumber \\
\end{eqnarray}
with
\begin{eqnarray}
\label{eq:scs2}
E_c^{MP2(OS)} &=& \sum_{ijab}\frac{(ia|jb)^2}{\epsilon_i+\epsilon_j-\epsilon_a-\epsilon_b}  \\
E_c^{MP2(SS)} &=& \sum_{ijab}\frac{(ia|jb) [(ia|jb)-(aj|bi)]} {\epsilon_i+\epsilon_j-\epsilon_a-\epsilon_b},
\end{eqnarray}
where $i,j$ denote occupied orbitals, $a,b$ denote virtual orbitals, 
$(ia|jb)$ denotes a two--electron integral in the Mulliken notation,
$\epsilon_i$ is the energy of the $i$--th orbital, and $c_{OS}$ and
$c_{SS}$ are the scaling coefficients for the opposite spin (OS) and
same spin (SS) correlation, respectively.
This approach was shown to yield quite improved results with respect
to the standard MP2, when a proper choice of the scaling parameters is
performed \cite{riley12,sosint07,onlyss07}. Alternatively, a good
performance was also obtained by limiting the method to consider only
the opposite spin part (i.e. setting $c_{SS}=0$ and optimizing only $c_{OS}$),
resulting in the scaled--opposite--spin MP2 (SOS--MP2) method \cite{jung04}, 
which allows, when properly implemented, to reduce the 
computational scaling to $O(N^4)$.
However, the choice of the optimal scaling parameters in SCS-- and SOS--MP2 is 
not trivial and different proposals have been made \cite{WCMS12}, either
based on theoretical considerations or on empirical fittings, showing 
that the ``best'' scaling parameters are somehow system and basis--set
dependent. This issue may be not particularly relevant for covalent bonds, 
where the considered binding energies and improvements over MP2 are much larger
that the inaccuracies due to a non--optimal choice of the parameters. It
becomes anyway relevant for noncovalently bonded complexes where the energies
to be computed are much smaller. 

For these latter cases, 
very good SCS parameters were proposed by
Distasio and Head--Gordon (SCS(MI)--MP2 method) 
\cite{sosint07} by optimizing the scaling coefficients against 
benchmark noncovalent interaction energy data. The same authors,
as well as Grant Hill and Platts in a separate paper,
also showed that accurate results can even be achieved by
just same--spin scaling (SSS--MP2; $c_{OS}=0$, $c_{SS}\approx$1.75) 
\cite{sosint07,onlyss07} methods. 
Additional investigations concerned the optimization of the
SOS--MP2 for long--range interaction by the introduction in the
SOS--MP2 method of a distance--dependent scaling coefficient and
a modified distance--dependent two--electron operator \cite{lochan05}.

The effort spent to optimize the scaling parameters and the fact that
the results of these works are not in an agreement with ``standard''
SCS--MP2 ($c_{OS}=6/5$, $c_{SS}=1/3$) or SOS--MP2 ($c_{OS}=1.3$) values nor
with the values suggested by theoretical arguments within wave 
function theory \cite{szabados06,fink10,WCMS12}, indicates the
difficulty to fix optimal values of the scaling for the proper
description of noncovalent interactions. Moreover, all these approaches
pay the prize of introducing a high level of empiricism.
Therefore, a simple non--empirical procedure to fix the
value of the various scaling parameters appears highly desirable.
\par In this paper we address this issue and we show a simple
 non--empirical scheme to fix the scaling factor in calculations of
the dissociation curve of noncovalent bonded complexes. 
 Here, non--empirical denotes the fact that scaling factors will
be fixed by direct  use of  information from high--level {\it ab initio} 
calculations and not from some empirical fitting procedure. 
Of course, some empiricism is still implied in the use of
 spin--resolved MP2 formulas.
Our approach focuses on scaled MP2 calculations
with a single parameter. To this end, we 
introduce the SOS(R)--MP2, 
SSS(R)--MP2, and scaled MP2 (S(R)--MP2, i.e.
with $c_{SS}=c_{OS}$) methods. 
 In acronym (R) stands
for ``calculated from Reference non--empirical values'', 
to avoid confusion and distinguish from empirically
scaled versions of standard SCS--MP2 methods.
 Our procedure is based on
the observation that there exist a well defined proportionality between
scaled--(same/opposite)--MP2 energies and the correlation energy
computed by high--level methods (e.g. CCSD(T)), which is almost 
independent on the intermolecular distance. 
Thus, the scaling parameter can be fixed 
once by using the information from only one expensive high--level calculations
and successively the whole PES can be computed with high accuracy by 
performing only relatively cheap scaled--(same/opposite)--MP2 calculations.
In particular, in this way the efficiency of the SOS--MP2 (SOS(R)--MP2) 
method ($O(N^4)$ scaling  \cite{jung04})  can be 
fully exploited for large scale explorations of dissociation 
potential energy surfaces (PES) without introducing
empirical parameters and achieving almost the CCSD(T) accuracy. 

We acknowledge that a similar approach was already recently used in
Ref. \cite{usvayat2013}, to compute accurate interaction energies of an Ar monolayer
adsorbed on an MgO substrate. However, in Ref. \cite{usvayat2013} only local MP2
calculations were performed while spin-resolved MP2 calculations, which are the main topic of
the present work, were not considered. 
In addition, in the present work, a systematic assessment of 
the non-empirical procedure for fixing the scaling parameters is carried out for 
different systems and interactions.

\section{ Method}\label{Sec:Theoscscoeff}

\begin{figure}
\includegraphics[width=\columnwidth]{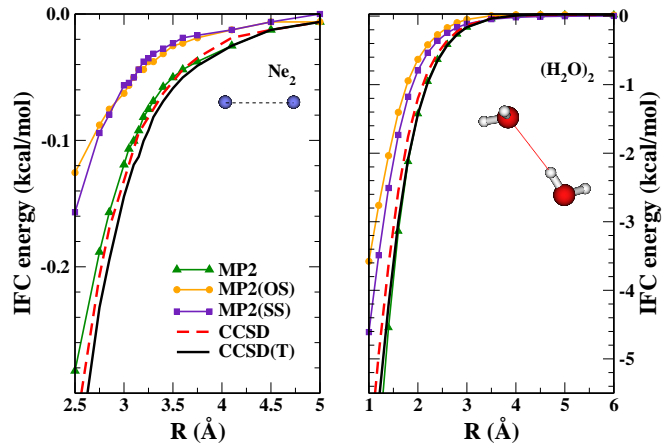}
\caption{\label{fig_ifc} Inter--fragment correlation (IFC) energy of Ne$_2$ and (H$_2$O)$_2$ as computed with different methods.}
\end{figure}

The main quantity of interest in the present paper is the
inter--fragment correlation (IFC) energy, defined as:
\begin{equation}\label{corrintenerX}
E_{IFC}^{X} = E_{c,AB}^{X} - E_{c,A}^{X}-E_{c,B}^{X},
\end{equation}
where in our calculations 
$X$=CCSD(T), CCSD, MP2, SOS(R)--MP2, SSS(R)--MP2 or S(R)--MP2. The IFC is the difference
between the correlation energy ($E_{c,AB}$) of a complex ($AB$) and that of its 
constituting fragments (A and B), computed with a method $X$.
The IFC for two exemplary cases (H$_2$O dimer and Ne$_2$) is reported in 
Fig. \ref{fig_ifc} as a function of the intermolecular distance ($R$) 
for different methods.
The IFC is always negative and its absolute value rapidly decreases with $R$.

\begin{figure}
\includegraphics[width=\columnwidth]{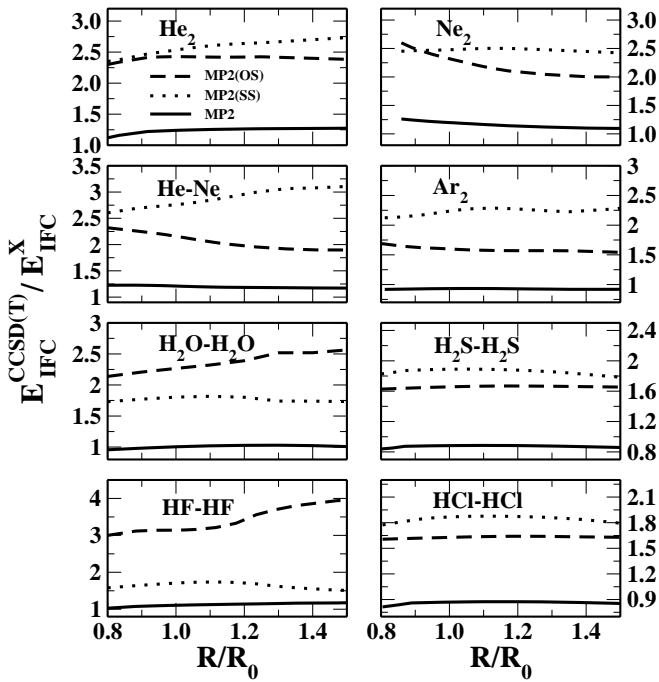}
\caption{\label{fig_ratio}Ratio between different spin--resolved MP2 IFC energies and CCSD(T) ones, 
for various systems at several distances. $R_0$ denotes the equilibrium distance 
of each complex. $X$ = MP2 [continuous line], MP2(OS) [dashed line], and MP2(SS) [dotted line].}
\end{figure}

It can be observed 
that for noncovalent interactions, with good approximation, the IFC 
energies computed at the MP2 (or equivalently MP2(SS) or MP2(OS)) level 
and at the CCSD(T) level are proportional to each other over a wide 
range of inter--fragment distances $R$.
This proportionality can be verified by direct comparison of the 
IFC energies calculated at different levels of theory for
various noncovalent interacting complexes
(see Fig. \ref{fig_ratio}). 
 For  MP2 the proportionality is very well satisfied for all systems 
and distances. MP2(SS) and MP2(OS) show instead some larger deviations 
at higher distances. However such deviations are
not very relevant, because the IFC energy is rapidly approaching zero 
for large distances. In fact, as shown in Fig. \ref{fig_ifc}, for $R/R_0\gtrsim$1.5 the
IFC is already close or below  subchemical accuracy.

The  proportionality of the (spin--resolved) MP2 and coupled 
cluster energies can be rationalized considering that in 
noncovalent complexes 
the variation of IFC with distance is
mainly due to relaxation of 
intramolecular contributions,
thus both MP2 and CCSD(T) IFC energies display a very similar 
behavior with $R$ and their ratio stays
almost constant.

This proportionality can be used to define
single--parameter scaled (spin--resolved) 
MP2 approaches suitable for the description of
the dissociation of noncovalent complexes.
To do that, we define
\begin{eqnarray}
\label{e6}
E_{IFC}^\mathrm{S(R)-MP2} & = & c_{S}E_{IFC}^\mathrm{MP2} \\
\label{e7}
E_{IFC}^\mathrm{SOS(R)-MP2} & = & c_{OS}E_{IFC}^\mathrm{MP2(OS)} \\
\label{e8}
E_{IFC}^\mathrm{SSS(R)-MP2} & = & c_{SS}E_{IFC}^\mathrm{MP2(SS)}\ ,
\end{eqnarray}
with
\begin{eqnarray}
\label{e9}
c_{S}  &=& \frac{E_{IFC}^\mathrm{CCSD(T)}(\tilde{R})}{E_{IFC}^{MP2}     (\tilde{R})} \\
\label{e10}
c_{OS} &=& \frac{E_{IFC}^\mathrm{CCSD(T)}(\tilde{R})}{E_{IFC}^{MP2(OS)}(\tilde{R})}\\
\label{e11}
c_{SS} &=& \frac{E_{IFC}^\mathrm{CCSD(T)}(\tilde{R})}{E_{IFC}^{MP2(SS)}(\tilde{R})}
\end{eqnarray}
where $\tilde{R}$ is some reference distance.
With this definition we fix (at the cost of a single
expensive CCSD(T) calculation) the proportionality between (spin--resolved) MP2
and CCSD(T) results, by simply imposing that at some point,
the two IFC energies, or analogously, the respective  total binding energies ($E_b$) 
are the same. 

The choice of the reference point $\tilde{R}$ in Eqs. (\ref{e9})--(\ref{e11}) is
not a major concern, since
the value of $c$ is almost independent on it (see Fig. \ref{fig_ratio} and
subsection \ref{sect_dist}).
In this work we have chosen for the the reference point 
the equilibrium distance $R_0$, 
computed at the CCSD(T) level. 
This choice is not mandatory, nevertheless it
appears the best compromise between the need to avoid
too short distances 
(where the proportionality relation may be less accurate)
and the necessity to avoid the use of too small energies 
(as would result for large $R$ values) in the ratio of Eqs.
(\ref{e9})--(\ref{e11}) to minimize numerical noise(see subsection \ref{sect_dist} for a
further discussion).

Each of the equations (\ref{e9}), (\ref{e10}), (\ref{e11}) provides a simple
non--empirical procedure for fixing the
(system--dependent) scaling factor in one--parameter (SCS--)MP2 calculations
when dissociation energies of noncovalent systems are of interest.
Thus, a whole dissociation curve, comprising many single--point results,
can be simulated by performing only a single CCSD(T) calculation and
without the introduction of any empirical parameter.

The main question is if the accuracy of the such procedure is high enough.
In section \ref{sec:res} we will show that deviations from the CCSD(T) reference results
are indeed well below subchemical accuracy (i.e. 0.1 kcal/mol).

\subsection{Computational details}
\label{sec_compdet}
To test our scaling approach we considered a representative set of small 
noncovalently interacting systems: He$_2$, Ne$_2$, He--Ne, Ar$_2$ (dispersion 
interaction), (H$_2$O)$_2$, (HF)$_2$ (hydrogen bond), (H$_2$S)$_2$,
(HCl)$_2$ (dipole--dipole interaction).  
Additionally, benzene--HCN,  the stacked benzene dimer, and Be$_2$ have
been also considered, as  particular cases. The former  two are in fact relatively 
large systems  with the first one also displaying 
a mixed electrostatic--dispersion character,
which changes with the bond distance \cite{hobza:2010:JCTC}.
The latter is a system where MP2 and CCSD completely fail  even qualitatively 
\cite{schmidt:2010:be2}.
The sizes of the molecules included in our  test set were limited by the need 
to compute in all cases the full CCSD(T) dissociation curves for reference purposes,
except for benzene--HCN  and the stacked benzene dimer in which case the 
reference data were taken from Ref.
\cite{hobza:2010:JCTC}.

 Calculations were performed with the ACES II \cite{acesii} and 
TURBOMOLE \cite{TURBOMOLE} program packages.
For all systems an aug--cc--pVQZ basis set \cite{dunn89,woon94,woon93} was used, 
except for Ne$_2$ (uncontracted aug--cc--pVTZ), Be$_2$ (cc--pV5Z), and 
Ar$_2$ (aug--cc--pV5Z).

To assess the performance of the different methods to reproduce the full dissociation curves
we considered the mean absolute error:
\begin{equation}\label{emae}
\mathrm{MAE}(X) = \frac{1}{R_{max}-R_{0}}\int_{R_{0}}^{R_{max}}|\Delta E^X(R)|dR\ ,
\end{equation}
where $X$=CCSD, MP2, SOS(R)--MP2, ..., and 
\begin{equation}\label{edelta}
\Delta E^X(R) \equiv E_{IFC}^{X}(R) - E_{IFC}^{CCSD(T)}(R)\ .
\end{equation}
Note that because all the methods considered here use the Hartree--Fock
exchange, the difference between the IFC energies of different methods in Eq.
(\ref{edelta}) 
are exactly the same as the differences in the total interaction energies. 
The MAE$(X)$ defined in Eq. (\ref{emae}) measures the average deviation of 
the results from the reference CCSD(T) data over a given interval.
In our work we decided to remove from this evaluation the smallest values of
$R$, because our approach might be not fully justified at such small 
inter--fragment distances. 
Hence, the lower bound for the integration was fixed at $R_0$. The
upper bound of the integral was fixed instead to the value of $R$ beyond
which the CCSD(T) IFC energy is lower than 10$^{-4}$ Hartree, in order to keep
the normalization factor finite in Eq. (\ref{emae}).

Finally, we note that in all calculations no correction for the basis
 set superposition error (BSSE) was included,
for computational simplicity and because it is readsorbed in the scaling 
coefficient.
For the water dimer, e.g., we check that, including a BSSE correction, 
the MAE of Eq. (\ref{emae}) is 36.7, 36.6 , 5.5 $10^{-3}$ kcal/mol, 
for  SOS(R)-MP2, SSS(R)-MP2 and S(R)-MP2 respectively. These values
are in good agreement with the ones (without BSSE) reported in Tab. \ref{tab2}.

\section{Results}\label{sec:res}
The values of the $c_{S}$, $c_{OS}$, and $c_{SS}$ coefficients, obtained
by applying Eqs. (\ref{e9}), (\ref{e10}), and (\ref{e11}) 
to the systems considered in this work, are 
reported in Table \ref{tablecos}, together with the reference
inter--fragment separation $R_0$.
\begin{table}
\small
\begin{center}
\caption{\label{tablecos} Values of the  $c_{OS}$,  $c_{SS}$, and  $c_{S}$
coefficients  obtained via Eqs. (\ref{e9}), (\ref{e10}), (\ref{e11}) calculated at
distance $R_0$ for all the systems presented in this paper.}
\begin{ruledtabular}
\begin{tabular}[c]{lccccccc}
system          & $\;$&   $ c_{OS} $  &  $c_{SS}$ &  $c_{S}$  &$\;$& $R_0$[\AA] \\
\hline
He$_2$          &&     2.43      & 2.53     &     1.24 &&   3.0  \\
He--Ne          &&     2.15      & 2.76     &     1.21 &&   3.0  \\
Ne$_2$          &&     2.34      & 2.47     &     1.20 &&   3.2  \\
Ar$_2$          &&     1.60      & 2.24     &     0.93 &&   3.7  \\
H$_2$S--H$_2$S  &&     1.66      & 1.89     &     0.88 &&   2.8  \\
HCl--HCl        &&     1.63      & 1.87     &     0.87 &&   2.5  \\
HF--HF          &&     3.14      & 1.71     &     1.11 &&   1.8  \\
H$_2$O--H$_2$O  &&     2.27      & 1.80     &     1.00 &&   2.0  \\
\end{tabular}
\end{ruledtabular}
\end{center}
\end{table}
Inspection of the table confirms that the optimal scaling coefficients
for (spin--resolved) MP2 calculations are significantly system dependent.
This dependence is also more pronounced for the SSS(R)--MP2 method, in which
MP2(SS) includes
only a small part of the total correlation contribution, while it is
much weaker for MP2, which can describe better the whole correlation effects.
It is interesting to note, in addition, that the variation of the 
scaling coefficients among different systems is not the same for the 
various methods. In fact, for example the $c_{OS}$ is maximum for HF--HF
and almost twice as big as the $c_{OS}$ for Ar$_2$, while
the value of $c_{SS}$ for HF--HF is about 20\% smaller than the 
Ar$_2$ one and it is maximum for He--Ne.
Therefore, no clear trend can be identified for the values of the 
scaling coefficients when different systems (and methods) are considered.

The results of Tab. \ref{tablecos} also indicate that, due to the variability
of the optimal scaling coefficients, global scaling factors can hardly 
be expected to achieve high accuracy for a broad range of applications.
In fact, even if most of the coefficients displayed in the table
agree reasonably well with the SOS(MI)--MP2 ($c_{OS(MI)}\approx1.8$) and 
SSS(MI)--MP2 ($c_{SS(MI)}\approx1.75$) values \cite{onlyss07,sosint07},
which were especially optimized for intermolecular interaction energies,
some remarkable differences appear. These reflect the peculiarities
of the correlation in some systems (e.g. same--spin correlation in He$_2$
or opposite--spin correlation in HF--HF) which cannot be captured
by ``average'' scaling procedures. Note finally, that all the $c_{OS}$
values are rather different from the conventional scaling coefficient of
1.3 proposed for quantum chemical applications \cite{jung04}.

\begin{figure}
\includegraphics[width=\columnwidth]{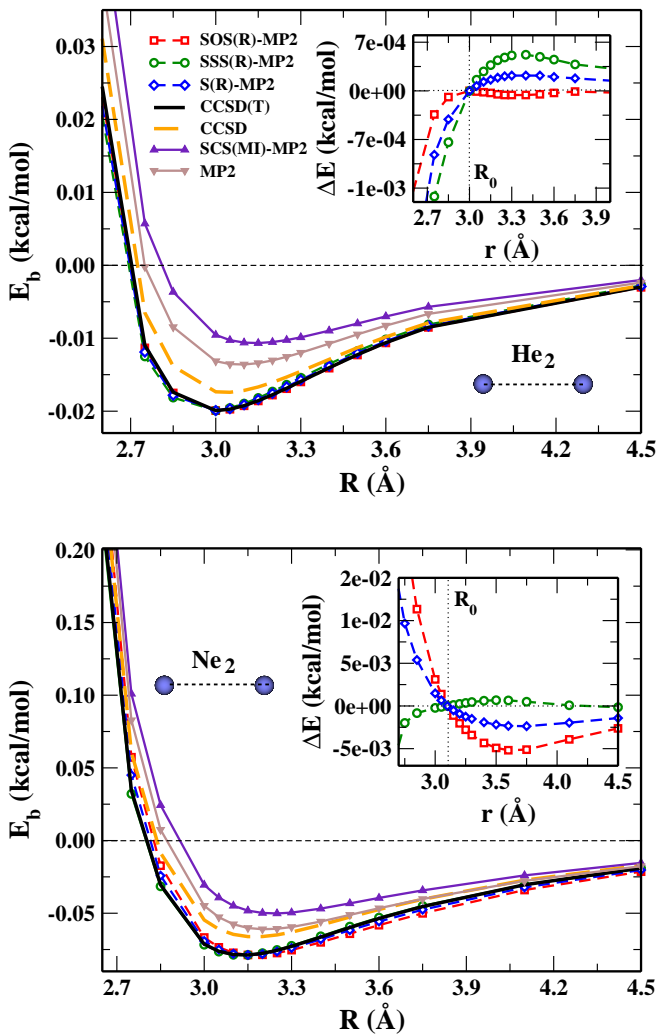}
\caption{\label{fig3_1} Binding energy of He$_2$ and Ne$_2$ as computed with 
different theoretical methods. In the insets we report the difference
 between S(R)--MP2, SSS(R)--MP2, SOS(R)--MP2 and CCSD(T) inter--fragment correlation energies
(see Eq. (\ref{edelta})). $R_0$ denotes the equilibrium distance.}
\end{figure}

To demonstrate the performance of the method we
report in Fig. \ref{fig3_1} the binding energies of two
dispersion dimers (He$_2$ and Ne$_2$) as computed with different 
methods.
The figure shows that the proposed scaled--(spin--resolved)--MP2
approaches yield very good results, giving binding energy curves which are
almost indistinguishable form the reference CCSD(T) ones over the whole range 
of inter--atomic distances considered. On the contrary, visible
differences with the reference are found by considering CCSD, MP2,
and SCS(MI)--MP2 results.
To have more insight into these results we show in the
insets of Fig. \ref{fig3_1} also the differences between our S(R)--MP2, SOS(R)--MP2, 
and SSS(R)--MP2 and the CCSD(T) reference ($\Delta E$, Eq. (\ref{edelta}))
at various distances.
By definition all our approaches coincide with CCSD(T) at the 
equilibrium (reference) distance $R_0$. Remarkably, however, the 
accuracy of the method is found to be very good at any distance,
in particular for $R > R_0$, where deviations from CCSD(T) values are 
vanishing small (of the order of 10$^{-3}$ kcal/mol). Such accuracy is 
not only well below subchemical accuracy but also largely sufficient to yield a 
highly accurate description of these dispersion dimers, as demonstrated by 
the comparison of the binding energy curves.

\begin{figure}
\includegraphics[width=\columnwidth]{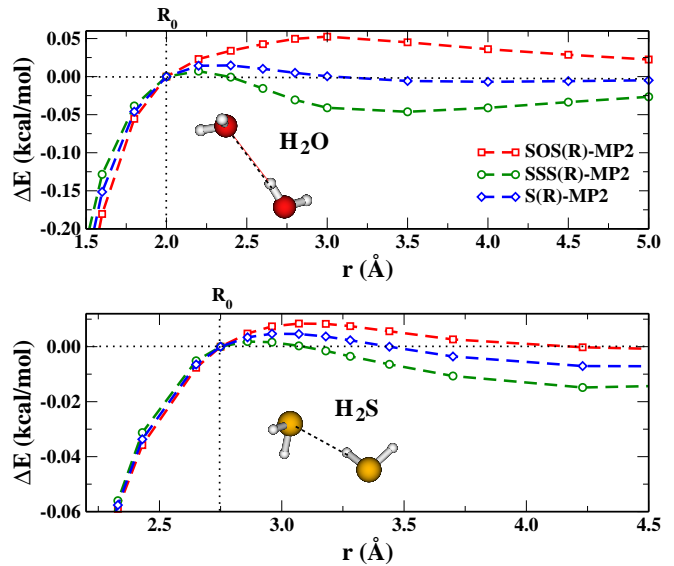}
\caption{\label{fig3_2} The difference ($\Delta E$, see Eq. (\ref{edelta})) between
S(R)--MP2,
SSS(R)--MP2, 
SOS(R)--MP2 and CCSD(T) inter--fragment correlation energies calculated for
(H$_2$O)$_2$ and (H$_2$S)$_2$.
 $R_0$ denotes the equilibrium distance.}
\end{figure}

As further examples Fig. \ref{fig3_2} reports the $\Delta E$ values
also for one hydrogen--bond complex (H$_2$O dimer) and one
dipole--dipole complex (H$_2$S dimer). In this case the binding curves
are instead not reported because due to the relatively large value of 
$E_b$ the differences between the different methods are not easily visible
on that scale.
Inspection of the figure shows again that the present scaled approaches
provide very accurate IFC energies over a broad range of intermolecular
distances, with discrepancies from the reference well below the
subchemical accuracy.
The analysis of these results, together with those of Figs. \ref{fig_ratio}
and \ref{fig3_1}, indicates that the S(R)--MP2 approach is overall
slightly more accurate than SOS(R)--MP2 and SSS(R)--MP2, but all yield highly accurate 
results. For short distances, however, the methods necessarily face some 
limitations. In fact, as we mentioned above, at short inter--system separations
the proportionality relation between (spin--resolved) MP2 and CCSD(T) may
deteriorate due to the  raising importance of
intermolecular correlation (that is, the direct interaction of 
electrons and pairs ``localized'' on different systems, starts to be relevant). 
Moreover, even in absence of this phenomenon, i.e. for 
systems and distances $R<R_0$ where 
Fig. \ref{fig_ratio} suggests that the proportionality holds, the 
values of $\Delta E$ may be expected to rapidly increase (in absolute
values) as the IFC energy is very large for $R<R_0$, so that 
small deviations are largely magnified.

To have a more quantitative and global assessment of 
the quality of the scaling coefficients calculated within
our simple non--empirical scaling scheme we report in Tab. \ref{tab2}
the mean absolute error, computed via Eq. (\ref{emae}), for several test
systems and various methods, including the spin--component scaled methods optimized for
intermolecular interactions \cite{sosint07}, the conventional SOS--MP2
with $c_{OS}=1.3$, the unscaled CCSD and empirical $1/R^6$, $1/R^5$, and
Lennard--Jones (LJ) potentials (fixed by fitting to CCSD(T) at $R_0$).
 Moreover, for comparison, we report the MAEs obtained by the popular PBE-D3
density functional (PBE exchange-correlation functional \cite{perd96a}
supplemented with the D3 empirical dispersion correction \cite{Grimme2010JCP};
 note that
the PBE functional is one of the best generalized gradient approximations for
non--covalent interactions \cite{Burns2001JCP,Fabiano:2011:JCTC}).
For DFT calculation the MAEs were computed by considering $E_b$ in place of $E_{IFC}$ into Eq. (\ref{edelta}). 

\begin{table*}
\small
\begin{center}
\caption{\label{tab2} Mean absolute error, as defined in Eq. (\ref{emae}),
 for different methods. Binding energies (E$_b$) are in kcal/mol, 
other results are in 10$^{-3}$ kcal/mol. RMAE (Eq. (\ref{RMAE}))
 is also reported in the last line.}
\begin{ruledtabular}
\begin{tabular}{lrrrrrrrrrrrrrrrrr}
System          & E$_b$ & SOS(R) &SSS(R) & S(R) & SCS(MI) & SOS(MI) & SSS(MI) &
SOS$^a$ & $1/R^6$ & $1/R^5$ & LJ & PBE-D3$^b$ & CCSD \\
\cline{3-9}
 & & \multicolumn{7}{c}{--MP2} & & & & & \\
\hline
He--He          & 0.02 &   0.0 &   0.3 &   0.1 &   3.0 &   2.4 &   3.2 &   4.5 &   0.3 &   2.0 &   5.3 &  1.3 &  0.7 \\
He--Ne          & 0.05 &   2.0 &   1.4 &   0.5 &  11.5 &   3.7 &  13.2 &  12.2 &   1.0 &   4.1 &  16.6 &  1.9 &  2.8 \\
Ne--Ne          & 0.08 &   3.0 &   0.3 &   1.5 &   9.6 &   5.8 &  10.4 &  14.3 &   1.7 &   4.0 &  15.2 &  6.5 &  4.3 \\
Ar--Ar          & 0.56 &   3.8 &   3.6 &   1.0 &  40.9 &  38.3 &  57.5 &  45.0 &  10.6 &  51.0 & 136.9 & 95.4 & 24.9 \\
H$_2$S--H$_2$S  & 1.97 &   3.3 &   9.0 &   4.5 &  23.9 &  61.7 &  40.3 & 151.2 & 123.3 &  29.6 & 167.6 & 160.3 & 110.1 \\
HCl--HCl        & 2.41 &   2.7 &   9.2 &   5.3 &  17.6 &  65.0 &  33.2 & 123.9 & 116.1 &  29.0 & 139.2 & 133.7 & 89.3 \\
HF--HF          & 4.88 &  78.4 &  36.7 &   5.4 &  29.2 & 112.3 &  40.1 & 125.3 &  54.1 &  51.1 &  74.1 &  169.8 & 30.3 \\
H$_2$O--H$_2$O  & 5.22 &  31.9 &  28.5 &   5.9 &  25.1 &  57.5 &  29.6 &  86.2 &  48.7 &  41.9 &  62.9 &  101.9 & 29.8 \\
                &       &       &       &       &       &       &       &       &       &       &       &       \\
RMAE            &       & 1.37\%& 0.93\%& 0.53\%& 7.54\%& 5.34\%& 8.81\%& 11.23\%& 2.58\%& 4.61\%& 15.02\% & 6.82\% & 3.68\% \\
\end{tabular}
\end{ruledtabular}
\end{center}
\begin{flushleft}
a) Conventional SOS--MP2 with $c_{os}=1.3$.\\
b) These data were obtained by substituting $E_{IFC}$ with $E_b$ in
Eq. (\ref{edelta}).
\end{flushleft}
\end{table*}
The values in the table are all very small, and are reported in unit of 10$^{-3}$ kcal/mol.
The first column reports the binding energy computed at the CCSD(T) level: 
the considered systems span a wide range of intermolecular forces.
In the last line also the relative mean absolute error (RMAE) 
is reported, which is computed as:
\begin{equation} \label{RMAE}
\mathrm{RMAE}(X) = \frac{1}{N}\sum_i^N\frac{\mathrm{MAE}_i(X)}{E_b[i]}\, 
\end{equation}
where the sum runs over all the $N$ systems  and $E_b[i]$ is the binding
energy of the $i$-th complex. 
The RMAE allows a fair global assessment of all the results.

The results of Tab. \ref{tab2} clearly show that the MP2--based methods
using the here proposed non--empirical scaling can reproduce the reference
CCSD(T) results with an accuracy  below $1.5\%$ in all systems 
(with a maximum MAE of  0.078 kcal/mol),
with an accuracy two--three times better than the CCSD method (RMAE=3.68\%). 
In particular, the present S(R)--MP2 has the lowest RMAE (0.53\%) confirming 
its very high accuracy
(the S(R)--MP2 MAE is always extremely small, lower than 0.006 kcal/mol).
On the other hand, SCS(MI)--, SOS(MI)--, and SSS(MI)--MP2, despite 
performing very well in absolute terms
(the deviation from CCSD(T) ranges from 0.003 to 0.1 kcal/mol)
yield significantly worse results, with RMAE larger than $5\%$.

A relatively small RMAE is obtained by the
empirical $1/R^6$ fit. This result however depends on the fact that
in our test set simple dispersion dimers are predominant. The empirical 
potential works however much worse for other cases.
In any case the RMAE of the $1/R^6$ fit is five times 
worse than that of the S(R)--MP2 approach.

 Finally, the PBE-D3 approach works fairly well,
yielding a RMAE comparable to that of the SOS(MI)-- and SCS(MI)--MP2
methods, but in any case more than five times larger than
all the methods using the here proposed
non--empirical scaling procedure. In addition, we note that
the PBE-D3 method yields very good results for dispersion
dimers, but not so good accuracy for hydrogen-bond and
dipole-dipole complexes.

\subsection{Role of the reference distance}
\label{sect_dist}
In this subsection we shortly analyze the role of the
reference distance used to compute the scaling coefficients
in Eqs. (\ref{e9}), (\ref{e10}), and (\ref{e11}). 
To this end we report in Tab.  \ref{tab3} the relative variation of the scaling
coefficients and the MAEs (Eq. (\ref{emae})) when the reference distance is changed
from the equilibrium value $R_0$ to $\tilde{R}>R_0$.

\begin{table*}
\small
\begin{center}
\caption{\label{tab3} Relative values (with respect to equilibrium geometry) of the scaling coefficients (obtained via Eq. (\ref{e9})) and mean absolute errors
 (Eq. (\ref{emae})) calculated at non equilibrium distances ($\tilde{R}>R_0$) for all the systems presented in this paper.}
\begin{ruledtabular}
\begin{tabular}[c]{lccccccc}
                &                   &                        &                  &                & \multicolumn{3}{c}{Mean absolute error} \\
\cline{6-8}
System  & $\tilde{R}$ &   $ c_{OS} $  &  $c_{SS}$ &  $c_{S}$ & SOS(R)--MP2 &
SSS(R)--MP2 & S(R)--MP2 \\
 \hline
He$_2$         & 1.1 & 1.00 & 1.00 & 1.00 & 1.00 & 0.67 & 1.00 \\
Ne$_2$         & 1.1 & 0.95 & 1.01 & 0.98 & 0.60 & 0.67 & 0.67 \\
He--Ne         & 1.1 & 0.95 & 1.04 & 0.98 & 0.55 & 0.71 & 0.40 \\
Ar$_2$         & 1.1 & 0.99 & 1.02 & 1.00 & 0.47 & 0.64 & 1.10 \\
H$_2$O--H$_2$O & 1.3 & 1.11 & 0.97 & 1.03 & 0.96 & 1.06 & 0.90 \\
HF--HF         & 1.3 & 1.13 & 0.99 & 1.04 & 1.17 & 1.01 & 0.85 \\
H$_2$S--H$_2$S & 1.2 & 1.01 & 1.00 & 1.00 & 0.67 & 0.99 & 1.07 \\
HCl--HCl       & 1.2 & 1.01 & 1.00 & 1.00 & 0.85 & 1.00 & 1.00 \\
\end{tabular}
\end{ruledtabular}
\end{center}
\end{table*}

The tabulated results show that scaling coefficients change 
very little, as can be already inferred from Fig. \ref{fig_ratio}.
This provides a validation of the working hypothesis
and indicates the
robustness of our simple non--empirical scaling procedure.
By inspecting the relative MAEs reported in Tab.
\ref{tab3} 
we see that the accuracy of the 
methods is not only preserved but in many cases is also improved.
 The improvement is anyway rather small and can be
considered to lay within the numerical uncertainty related
to the definition of the MAE (Eq. (\ref{emae})).

This result is important because it remarks the 
possibility to utilize the present non--empirical
scaling procedure also in all those cases where the
equilibrium distance is not known, by simply
performing a single CCSD(T) calculations at any 
``reasonable'' bond distance.

\subsection{Benzene complexes} 
In this subsection we consider the application of our new 
scaling method 
to larger systems, namely the benzene--HCN  T--shaped complex
 and the stacked benzene dimer.

\begin{figure}
\includegraphics[width=\columnwidth]{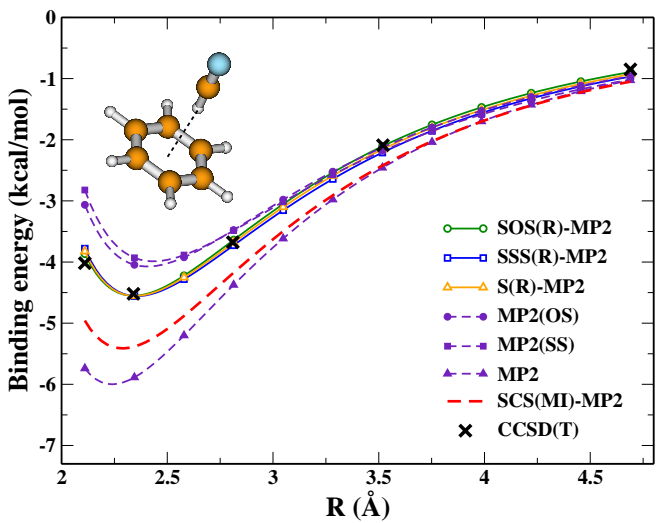}
\caption{\label{fig_benz} Binding energy for the benzene--HCN complex as a function of the inter--atomic distance $R$, for different methods.}
\end{figure}

 The first of these systems is particularly interesting because the
interaction has a mixed character involving both
electrostatic and dispersion contributions in a variable
proportion along the dissociation curve \cite{hobza:2010:JCTC}.
Its dissociation curves computed at the (spin--resolved) MP2 level
as well as the corresponding scaled results are reported in Fig. 
\ref{fig_benz}. Reference CCSD(T) data from Ref.  
\cite{hobza:2010:JCTC} are also reported. 
The scaling coefficients have been computed at the equilibrium distance
of 2.34 \AA{} and are $c_{OS}=1.27$, $c_{SS}=1.35$, and $c_S=0.66$. 

Consistently with the results discussed in the main body of
section \ref{sec:res}, the non--empirical scaling methods all perform
remarkably well, substantially
improving over ``bare'' MP2, MP2(OS), and MP2(SS) calculations
 and especially over empirically scaling methods (e.g. SCS(MI)--MP2).
S(R)--MP2, SOS(R)--MP2 and SSS(R)--MP2
show in fact deviations from the reference data 
below 0.5 kcal/mol along the whole range of distances considered.

Similar results are found also by using scaling coefficients
obtained considering a different reference distance $\tilde{R}$.
For $\tilde{R}=2.81$\AA, indeed, we found $c_{OS}$=1.15, $c_{SS}$=1.14,
$c_S$=0.57, and essentially the same performance for the binding energy
(or equivalently the IFC energy).
In fact, the scaling coefficients computed at the new distance are in good
agreement, although slightly smaller, with the ones computed at 
the equilibrium distance. We remark that the small difference between
the coefficients computed at different reference distances shall, in this
special case, be partially traced back to 
the computational noise originating from the fact that the
reference CCSD(T) values were not computed with the same set 
up as the MP2 ones, but rather extracted from the total binding 
energies (HF+CCSD(T) at the CBS limit) reported in Ref.  \cite{hobza:2010:JCTC}.
Despite this small issue, the non--empirically scaled methods
display an impressive robustness. 

\begin{figure}
\includegraphics[width=\columnwidth]{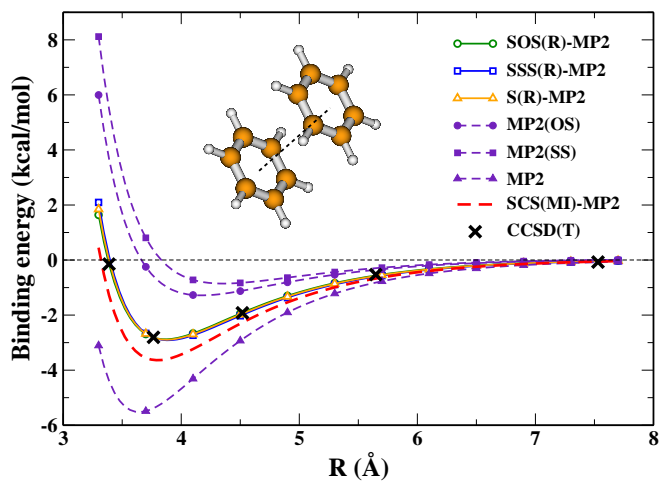}
\caption{\label{benzdimer} Binding energy for the stacked benzene dimer as a function of the inter--atomic distance $R$, for different methods.}
\end{figure}

 As a further test we consider a typical $\pi$-stacking complex.
In Fig. \ref{benzdimer} we thus report the dissociation curve of the stacked
benzene dimer as computed at the (spin--resolved) MP2 level and with the
various scaled approaches. Reference CCSD(T) data from Ref.  
\cite{hobza:2010:JCTC} are also reported.
The scaling coefficients have been computed at the equilibrium distance
of 3.765 \AA{} and are $c_{OS}=1.39$, $c_{SS}=1.66$, and $c_S=0.76$. 
Inspection of the figure shows that, similarly with the case of the benzene-HCN complex,
all the scaled methods perform remarkably well yielding deviations within 0.2 kcal/mol
from the reference CCSD(T) data. In particular, the non--empirical scaling procedure 
proposed here demonstrates to be able to correct well both the overbinding of the MP2 method
and the strong underbinding of the MP2(OS) and MP2(SS) approaches.

Overall, the dissociation curves computed with our scaling approach
agree very well with reference data and show that the method is very 
effective even for such ``large'' and ``difficult'' systems as the
present ones. We note that this performance is remarkable especially 
for SOS(R)-- and SSS(R)--MP2, because 
the MP2(OS) and MP2(SS) binding energies are almost degenerate,
so that the corresponding scaling coefficient are almost identical.
Thus, the same scaling is obtained for the two 
methods. However, this has only a minor effect on the SSS(R)--MP2
results at relatively small distances, while a good performance
is observed for bond distances larger than the equilibrium one.
This shows once more that the here presented scaling
procedure is rather robust in many different situations.

\subsection{Beryllium dimer}

\begin{figure}
\includegraphics[width=\columnwidth]{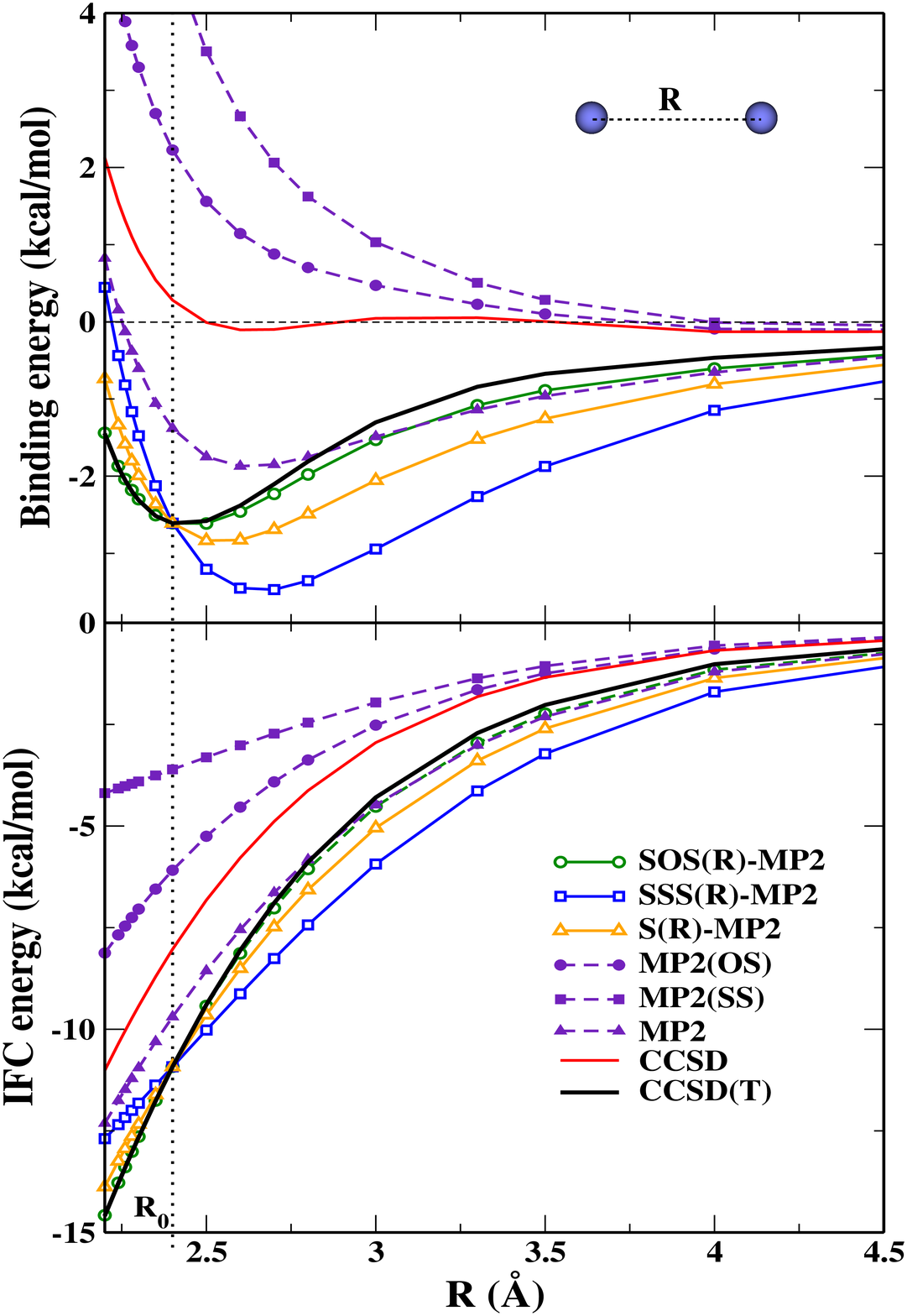}
\caption{\label{fig_be2} Binding energy and IFC energy for Be$_2$ as a function of the inter--atomic distance $R$, for different methods. $R_0$ denotes the equilibrium distance.}
\end{figure}

Finally, we consider in Fig. \ref{fig_be2} the Beryllium dimer,
which is a very difficult system where MP2 and CCSD fail to
give even an approximate quantitative description of the dissociation curve
\cite{schmidt:2010:be2}. For this system even CCSD(T)
shows some limitations \cite{schmidt:2010:be2}. However, 
this high-level method can at least capture most of the features of the dissociation curve.
Therefore, it will be used anyway as reference in the present example, although caution must be
paid to the quantitative analysis of the results.

In this case, the computed scaling coefficients are
$c_{OS}=1.79$, $c_{SS}=3.03$, and $c_{S}=1.13$. However,
an accurate description of the dissociation
curve is found only  using the SOS(R)--MP2 method.  
On the contrary, methods based on the same--spin correlation
fail quite evidently. This may be due to the fact that in this system,
where static correlation is important, the stretching of the bond is not only
promoting polarization and induction phenomena in each atom, but is also
changing the actual ``multireference'' description of the system,
influencing the same--spin correlation which includes antisymmetrized
integrals. 
This fact indicates that the use of MP2(SS) correlation energy 
in scaling procedures has strong limitations for this peculiar system. 
Due to this limitations
also the S(R)--MP2 method, which includes important contributions
of same--spin correlation, cannot be accurate.
Nevertheless, we remark that S(R)--MP2 and even SSS(R)--MP2 can compensate the
limitations of the underlying MP2(SS) energy relatively well,
thanks to appropriate values of the scaling factor
granted by Eqs. (\ref{e9}), (\ref{e10}), (\ref{e11}). However, the independence of the scaling factor 
from the reference separation is partially lost and only a limited portion
of the dissociation curve can be properly described when same--spin
correlation is considered.
\par The SOS(R)--MP2 method with coefficients calculated in our scaled
procedure provides for the Be$_2$ the best dissociation curve 
(as compared to CCSD(T))
significantly outperforming MP2, CCSD or SOS(MI)--MP2 methods.

\section{Conclusions}

We have proposed a simple non--empirical procedure which can be used to
calculate optimal scaling coefficients for (spin--resolved) MP2 calculations of the
dissociation of noncovalent complexes. 
The applicability of the proposed method was demonstrated for a series
of test systems, considering equilibrium and non--equilibrium reference distances,
as well as for a notoriously
difficult case as Be$_{2}$. The obtained results show that
the presented method works well in most cases, confirming the robustness of
our hypothesis.

The proposed method is especially attractive because it allows to
obtain a full dissociation curve of almost CCSD(T) quality
at the cost of just a single CCSD(T) calculation.
Moreover, it is conceivable to replace the costly 
CCSD(T) calculation with a focal--point analysis 
\cite{east;1993;focal,csaszar;1998;focal} (the $\Delta$CCSD(T) procedure) 
to obtain an estimate of the CCSD(T) correlation energy 
from a cheaper calculation. In this way the computational cost is 
much reduced, losing only little on accuracy.
Thus, future applications on large systems can be foreseen.
Moreover, it might be also possible to replace the CCSD(T)
reference with an alternative accurate treatment of the correlation,
e.g. the  CCSD[T] \cite{urban:1987:ccsd[t],hobza:2013:CCSD[T]} 
 or the FNO CCSD(T)
\cite{sherill:2013:FNO} 
 method, to optimize
the accuracy or the computational effort. 

Finally we remark that, although in this paper we focused only on
single--parameter scaled MP2 methods, the methodology presented
here can be extended in the future to treat the more
general case of spin--component--scaled (SCS) MP2 calculations
with two parameters ($c_{OS}$ and $c_{SS}$). To this end
Eqs. (\ref{e6})-(\ref{e8}) will be generalized to
\begin{equation}\label{2parifc}
E_{IFC}^\mathrm{CCSD(T)}(\tilde{R})=c_{OS} E_{IFC}^{MP2(OS)}(\tilde{R})+
c_{SS} E_{IFC}^{MP2(SS)}(\tilde{R})\; ,
\end{equation}
and an additional constraint will be needed for the scaling parameters (e.g.
fixing the sum or the ratio of the parameters from theoretical
considerations). 
We also have to stress that this approach can be  useful 
for the verification of the quality of SCS coefficients used in 
different SCS--MP2 methods.

\textbf{Acknowledgments}
This work was supported by the Polish Committee for Scientific
Research MNiSW under Grant no. N N204 560839 and by 
the ERC--StG FP7 Project DEDOM (no. 207441). We thank 
TURBOMOLE GmbH for providing the TURBOMOLE program package and M. 
Margarito for technical support.

\end{document}